\documentclass[pra,twocolumn,superscriptaddress,showpacs,amssymb]{revtex4}
\usepackage{graphicx}
\usepackage{epsfig} 
\usepackage{graphics}

\newcommand{\braket}[2]{\left \langle #1 | #2 \right \rangle}

\newcommand{\an}[1]{\hat{#1}}
\newcommand{\ad}[1]{\hat{#1}^{\dag}}
\newcommand{\be}{\begin{equation}}
\newcommand{\ee}{\end{equation}}
\newcommand{\bea}{\begin{eqnarray}}
\newcommand{\eea}{\end{eqnarray}}
\newcommand{\ket}[1]{\left | #1 \right \rangle}
\newcommand{\bra}[1]{\left \langle #1 \right |}

\begin{document}

\title{Dynamics of entanglement between two atomic samples with spontaneous scattering}

\author{Antonio Di Lisi} 
\email[E-mail:]{dilisi@sa.infn.it}
\author{Silvio De Siena}
\email[E-mail:]{desiena@sa.infn.it}
\author{Fabrizio Illuminati}
\email[E-mail:]{illuminati@sa.infn.it}
\affiliation{Dipartimento di Fisica ``E. R. Caianiello'', Universit\`a di Salerno,
INFM--UdR di Salerno, INFN Sezione di Napoli, Gruppo Collegato di Salerno,
I-84081 Baronissi (SA), Italy}

\date{October 30, 2003}

\begin{abstract}
We investigate the effects of spontaneous scattering on 
the evolution of entanglement of two atomic samples,
probed by phase shift measurements on optical beams interacting with
both samples. We develop a formalism of conditional quantum evolutions
and present a wave function analysis implemented in numerical simulations of the state vector 
dynamics. This method allows to track the evolution of entanglement and to
compare it with the predictions obtained when spontaneous scattering is neglected.
We provide numerical evidence that the interferometric scheme to entangle atomic
samples is only marginally affected by the presence of spontaneous scattering, and
should thus be robust even in more realistic situations. 
\end{abstract}


\maketitle

\section{Introduction}

In recent years much interest has been devoted to the study of quantum
\emph{entanglement}, one of the most profound consequences of quantum
mechanics and by now thought of as a fundamental resource of Nature,
of comparable importance to energy, information, entropy, or any
fundamental resource \cite{nielsen00}. Entanglement plays a crucial
role in many fundamental aspects of quantum mechanics, such as the quantum theory 
of measurement, decoherence, and quantum nonlocality. Furthermore, entanglement is
one of the key ingredients in quantum information and quantum computation theory, 
and their experimental implementations \cite{nielsen00}.

Many theoretical and experimental efforts have been recently devoted
to create and study entangled states of material particles 
by exploiting photon-atom interactions \cite{sackett,rausch}. 
The machinery of quantum non-demolition (QND) 
measurements provides an important tool toward this goal.
In particular it is possible to probe nondestructively the atomic state population
of two stable states of an atomic sample by phase-shift measurements on a field 
of radiation non-resonantly coupled with one of the atomic state 
\cite{kuz98,moelmer99,bouchoule02,kuz99,Kuz00}. 
The consequent modification of the state vector or density matrix of the sample
can be used to entangle pairs of
samples by performing joint phase-shift measurements on light propagating through 
both samples that provide information about the total occupancies of various states 
\cite{{Duan00}}. Recently, entangled states of 
two macroscopic samples of atoms were realized by QND
measurements of spin noise~\cite{JKP}. 

So far, most of the theoretical work has focused on the ``static'' 
entanglement properties, but very few results have been 
obtained on the dynamical properties of the entanglement of these systems, 
\emph{i.e.}, on how to realize the most efficient entangling dynamics
by using the interaction processes allowed by the physical set-ups. 
  
In order to better understand the dynamics of entanglement in coupled systems
of matter and radiation, a quantitative wave function analysis has been 
recently introduced \cite{noi3} to determine the 
entanglement created by measurements of total population on
separate atomic samples by means of optical phase shifts.
The authors of Ref.~\cite{noi3} considered a photon-atom 
interaction scheme in which dissipative effects, as
spontaneous scattering, were completely neglected. 
However, a proper inclusion of spontaneous scattering would provide
a first step toward an understanding 
of the decoherence effects affecting the system, and,
in view of achieving a practical control of the entangling protocol, it must be 
considered  in order to have a more realistic physical description of the process.   
    
In this paper we reformulate the interaction model presented in  Ref.~ \cite{noi3} so as to 
include spontaneous scattering in the wave function analysis, and provide a comparison 
of the results obtained from the two schemes of photon-atom interaction, with and
without spontaneous scattering. A byproduct of our study, that is not restricted
to the specific problem faced in the present work, is the formulation of  
a theory of conditional quantum evolution for generic photon-atom scattering
in systems with many atoms.

The paper is organized as follows. In Sec.~\ref{model} we present the interferometric set-up
and the detection scheme used to measure the field phase shift and to detect the scattered photons.
Moreover, we introduce the analytical model to determine how the wave function of the atomic samples
is modified by the photo-detection. We will distinguish between the case in which the photon
is scattered in the same or in a different mode with respect to the initial one. In Sec~\ref{CEP},
we analytically show how the two atomic clouds get entangled due to the photo-detection and give
the algorithm to implement the physical model in a numerical simulation. 
The results of two kinds of simulations are presented in Sec.~\ref{NS}. In the first one, we simulate
the consecutive measurement of two combinations of spin variables, 
showing how the entanglement of the two atomic samples evolve as the photo-detection occurs. 
In the second one, we present simulations where the atomic samples are subject to continuous spin rotations
during measurement, resulting in a different evolution of the entanglement. 
For both schemes of simulations we then compare the case in which the spontaneous scattering 
is considered with the case in which it is neglected. Finally, in Sec.~\ref{Conclusions} we
draw our conclusions.  
 
\section{The interferometric scheme and the wave function updating model}\label{model}

\begin{figure}[hbtp]
\begin{center}
\includegraphics[width=7.5cm]{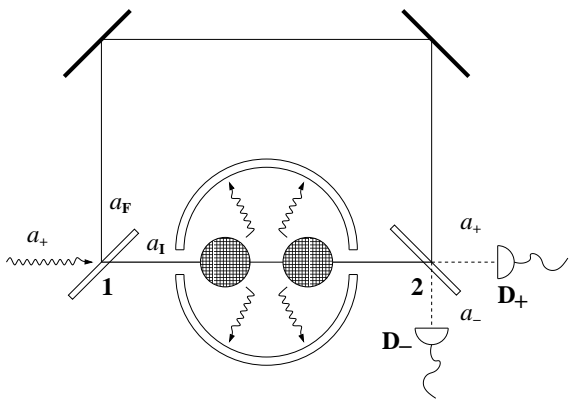}
\end{center} 
%%\vspace{2cm}
\caption{\small{ 
Atoms occupying the internal state $|a\rangle$ in the two samples interact 
with the light field which is incident from the left in the figure. 
The phase shift of the light field due to interaction with these atoms 
is registered by the different photo-currents in the two detectors.}}
\label{IS}
\end{figure}

Let us consider two atomic samples placed in one of the arms of an 
interferometric set-up used to measure the phase shift of the entering electromagnetic
field (Fig.~\ref{IS}).  
 
A photon impinging  the 50-50 beam splitter $1$ is ``separated'' in two 
components, denoted by $F$, which follows the upper \emph{free} path, and $I$, 
which can \emph{interact} with the two atomic samples in the lower arm.
$\an{a}_{F}$ and $\an{a}_{I}$ denote the annihilation operator modes of the two 
field components. 

\begin{figure}[hbtp]
\begin{center}
\includegraphics[height=4cm]{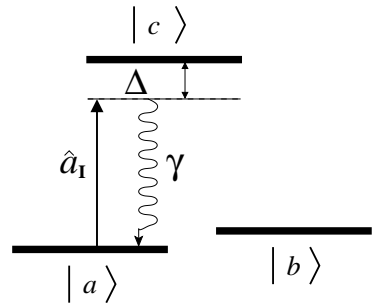}
\end{center} 
%%\vspace{2cm}
\caption{\small{ 
Level structure of the atoms. The states $\ket{a}$ and $\ket{b}$ are
stable states, $|a\rangle$ is coupled off-resonantly by the probe to 
the excited state $\ket{c}$. Here $\gamma$ is the spontaneous emission rate.}}

\label{LS}
\end{figure}

Each sample is composed of $N$ atoms, whose level structure consists of two
stable states, $\ket{a}$ and $\ket{b}$, and one excited state, $\ket{c}$ (Fig.~\ref{LS}).
States $\ket{a}$ and $\ket{c}$ are coupled off-resonantly by the electromagnetic field 
whose annihilation operator is $\an{a}_I$. 
Thus, a photon passing through the two atomic samples can be either absorbed 
and spontaneously re-emitted, or transmitted unchanged, depending on the atomic state 
populations. A second beam splitter $2$ recomposes the two original field modes, 
and the photo-currents induced in detector $D_+$ and $D_-$ allow to measure the 
field phase shift. This measurement modifies the state vector 
of the two atomic clouds that become entangled, and at the same time 
it allows to probe the entanglement evolution.

Due to the spontaneous scattering, not all the photons will be revealed by the detectors 
$D_+$ and $D_-$. Therefore, in order to get a more realistic description of the process,
it is important to evaluate the effects the scattered photons have 
on the atomic state vector evolution and on the entanglement dynamics. 
To this aim, we can imagine to detect the scattered photons  
by means of a spherical detector surrounding the two atomic clouds, and thus calculate  
how this detection, which corresponds to a spontaneous scattering event,
modifies the atomic wave function~(Fig.~\ref{IS}). Notice that in order to preserve the purity of
the atomic state we consider a spherical photodetector revealing not only the presence of
a scattered photon, but its (azimuthal) direction as well.

\subsection*{The reset operator}\label{RO}

In this and in the next sub-section we will develop a formalism to
account for the reset of the atomic sample wave function 
after the photons are detected. 

To this end, we first introduce a reset operator $\an{R}$, 
whose action on the atomic state 
$\ket{\Psi}$ yields a new non-normalized state $\ket{\Psi^{\prime}}$, affected
by the spontaneous scattering photo-detection, 
that must be normalized to the final atomic state.
The reset operator was first introduced in the context of the \emph{quantum jump} approach
to study the dynamical evolution of open quantum optical systems 
determined by continuous (gedanken) measurements on the radiated field
\cite{carmichael,moelmer92,moelmer93,Hegerfeldt93,beige98,beige99}. 
In deriving the analytic expression of the reset operator, we will consider
the formalism introduced by Hegerfeldt \cite{Hegerfeldt93} and properly adapt it
to our case study.  

Let us suppose that the photons are scattered one at a time by an atom belonging
indifferently to one of the two samples.
In the dipole and rotating wave approximations, the interaction
Hamiltonian reads
\bea
H_{I}(t)&=&\sum_{n}\sum_{\vec{k},\epsilon}
g_{\vec{k},\epsilon}\left(\ket{c}_{n}\!\bra{a}\ad{a}_{I}e^{i(\omega_0 - \omega_k)t}\right.\nonumber\\
&&\left. + \ket{a}_{n}\!\bra{c}\an{a}_{I}e^{-i(\omega_0 - \omega_k)t}\right)\;,\label{hamiltonian}
\eea  
where $n=1,2...,N$ goes over to the total number of atoms $N$, 
$\vec{k}$ and $\epsilon$ are the photon wave vector and polarization component respectively, 
$\omega_0$ is the energy difference between the atomic level $\ket{c}$ and $\ket{a}$,
$\omega_k$ is the field frequency, $g_{\vec{k},\epsilon}$ is the atom-photon interaction strength,
and $\an{a}_{I}$ is the annihilation operator for the field component interacting with the atoms.
Henceforth we will consider $\hbar=c=1$. Consequently 
$g_{\vec{k},\epsilon}$ can be defined as
\be\label{g}
g_{\vec{k},\epsilon}=
-\left(\frac{\omega_k}{2\varepsilon_0 L^3}\right)^{\frac{1}{2}}\vec{d}_{ac}\cdot\an{\epsilon}\;,
\ee
where $\vec{d}_{ac}$ is the transition dipole moment, the same for every atoms,
$\an{\epsilon}$ is the polarization vector and $L^3$ is the quantization volume.
We are considering fields whose wavelength is larger than the spatial dimension of the two samples.
Therefore, we can neglect not only the relative positions of the atoms, but even that of the two
atomic clouds, which are supposed to be placed in the origin of the coordinate system.
As the photon interacts only with one atom at the time, 
we will consider this case in the development of the calculations. 
The generalization to the simultaneous interaction with $N$ atoms is straightforward.  

During the interval $\Delta t \gg\frac{1}{\omega_0}$, 
the time evolution operator, up to second order perturbation theory, is

\bea\label{teo1}
U_I(t+\Delta t,t) &=& 1 - i\int_{t}^{t+\Delta t} dt^{\prime}H_{I}(t^{\prime})\nonumber\\
&&-\int_{t}^{t+\Delta t} dt^{\prime}\int_{t}^{t^{\prime}} dt^{\prime\prime}
H_{I}(t^{\prime})H_{I}(t^{\prime\prime})\;.\nonumber\\
&&
\eea  

Let $\ket{i} =\ket{\vec{k},\epsilon}\ket{\Psi}$ be the initial state of the 
system photon+atom; after the interaction has taken place,
it evolves in
%\be\label{fs0}
$\ket{\vec{k}^{\prime},\epsilon^{\prime}}\ket{\Psi_{k^\prime}}
=U_I(t+\Delta t,t)\ket{\vec{k},\epsilon}\ket{\Psi}$.
%\ee

Assuming that the photo-detection allows to discriminate only the direction of the photon wave vector,
the correct final state is:

\be\label{fs1}
\ket{f}=\sum_{k^{\prime},\epsilon^{\prime}}\ket{\vec{k}^{\prime},\epsilon^{\prime}}
\bra{k^{\prime},\epsilon^{\prime}}U_I(t+\Delta t,t)\ket{\vec{k},\epsilon}\ket{\Psi}.
\ee

From Eq.~(\ref{fs1}) it follows that
$\ket{\Psi_{k^\prime}}=
\bra{\vec{k}^{\prime},\epsilon^{\prime}}U_I(t+\Delta t,t)\ket{\vec{k},\epsilon}\ket{\Psi}$ 
and we can then introduce the reset operator 
\be\label{ro1}
\an{R}_{k^\prime}=\bra{\vec{k}^{\prime},\epsilon^{\prime}}U_I(t+\Delta t,t)\ket{\vec{k},\epsilon},
\ee
acting on the atomic state and, once the photon is detected, yielding the updated atomic 
state vector $\ket{\Psi_{k^\prime}}$.

In our case, the term of first order in Eq.~(\ref{teo1}) vanishes, and we have

$$
\!\!\!\!\!\!\!\!\!\!\!\!\!\!\!\!\!\!\!\!\!\!
\bra{\vec{k}^{\prime},\epsilon^{\prime}}U_I(t+\Delta t,t)\ket{\vec{k},\epsilon}= 
\delta_{k,k^{\prime}}\delta_{\epsilon,\epsilon^{\prime}} +
$$
\be 
\;\;-\int_{t}^{t+\Delta t} dt^{\prime}\int_{t}^{t^{\prime}} dt^{\prime\prime}
\bra{\vec{k}^{\prime},\epsilon^{\prime}}H_{I}(t^{\prime})H_{I}(t^{\prime\prime})\ket{\vec{k},\epsilon}.
\ee
Working out explicitly the second order term 
$\bra{\vec{k}^{\prime},\epsilon^{\prime}}
H_{I}(t^{\prime})H_{I}(t^{\prime\prime})\ket{\vec{k},\epsilon}$, 
we finally obtain
$$
\bra{\vec{k}^{\prime},\epsilon^{\prime}}U_I(t+\Delta t,t)\ket{\vec{k},\epsilon}=
\delta_{k,k^{\prime}}\delta_{\epsilon,\epsilon^{\prime}}
- \; g_{\vec{k},\epsilon}g^{*}_{\vec{k}^{\prime},\epsilon^{\prime}}
%\;+\;\;\;\;\;\;\;\;
%\;\;\;\;\;\;\;\;\;\;\;\;\;\;\;\;\;\;\;\;\;\;\;\;\;\;\;\;\;\;\;\;\;\;\;\;\;\;\;\;
%\;\;\;\;\;\;\;\;
$$
$$
\times\left[\int_{t}^{t+\Delta t} dt^{\prime}\int_{t}^{t^{\prime}} dt^{\prime\prime}
e^{i(\omega_0-\omega_{k^{\prime}})t^{\prime}}e^{-i(\omega_0-\omega_{k^{\prime}})t^{\prime\prime}}
\ket{c}\bra{c}\right.
$$
\be
+\left.\int_{t}^{t+\Delta t} dt^{\prime}\int_{t}^{t^{\prime}} dt^{\prime\prime}
e^{-i(\omega_0-\omega_{k^{\prime}})t^{\prime}}e^{i(\omega_0-\omega_{k^{\prime}})t^{\prime\prime}}
\ket{a}\bra{a}\right]\;.\;\;\;\;\;\;\;\;\;\;\;\;\;\;\label{ro2}
\ee 

As we are considering atoms initially in their stable states $\ket{a}$ and $\ket{b}$,  
the first integral in Eq.~(\ref{ro2}) acting on these states is always zero 
($\braket{c}{a}=\braket{c}{b}=0$). Thus, we are interesting only in the second integral.
%$\int_{t}^{t+\Delta t} dt^{\prime}\int_{t}^{t^{\prime}} dt^{\prime\prime}
%e^{-i(\omega_0-\omega_{k^{\prime}})t^{\prime}}
%e^{i(\omega_0-\omega_{k^{\prime}})t^{\prime\prime}}
%\ket{a}\bra{a}$.
The calculation is now similar to that of the transition amplitude for resonant scattering. 
Performing the substitution $t=-\frac{\tau}{2}$, $t+\Delta t=\frac{\tau}{2}$ $\Rightarrow$
$\Delta t=\tau$, and adapting the method used in Ref.~\cite{cohen92} for
the case of resonant scattering, we have
$$
\bra{\vec{k}^{\prime},\epsilon^{\prime}}U_I(t+\Delta t,t)\ket{\vec{k},\epsilon}=
\delta_{k,k^{\prime}}\delta_{\epsilon,\epsilon^{\prime}}+\;\;\;\;\;\;\;\;\;\;\;\;\;
$$
\be 
-2\pi i\frac{g_{\vec{k},\epsilon}g^{*}_{\vec{k}^{\prime},\epsilon^{\prime}}}{\Delta-i\frac{\gamma}{2}}
\delta^{\tau}(\omega_{k^{\prime}}-\omega_k)\ket{a}\bra{a}\label{ro3}, 
\ee 
where $\Delta=\omega_k-\omega_0$ is the detuning between the incident field and the atomic 
transition energy, $\gamma$ is the spontaneous emission rate of the transition 
$\ket{c}\rightarrow\ket{a}$, and the delta function $\delta^{\tau}$, 
expressing conservation of energy, reads $\delta^{\tau}(\omega-\omega^{\prime})=
\int_{-\frac{\tau}{2}}^{\frac{\tau}{2}}d\tau^{\prime}e^{i(\omega-\omega^{\prime})\tau^{\prime}}\nonumber
= \sin[(\omega-\omega^{\prime})] \tau / 2\pi(\omega-\omega^{\prime})$.
%\bea\label{dt1}
%\delta^{\tau}(\omega-\omega^{\prime})&=&
%\int_{-\frac{\tau}{2}}^{\frac{\tau}{2}}d\tau^{\prime}e^{i(\omega-\omega^{\prime})\tau^{\prime}}\nonumber\\
%&=&\frac{1}{\pi}\frac{\sin[(\omega-\omega^{\prime})]\frac{\tau}{2}}{\omega-\omega^{\prime}}.
%\eea
%It is worth to mention the following properties of $\delta^{\tau}$, widely used later on: 
%\bea
%\lim_{\tau\rightarrow\infty}\delta^{\tau}(\omega-\omega^{\prime})&=&\delta(\omega-\omega^{\prime}),
%\label{dt2}\\ 
%\omega\approx\omega^{\prime}&\Rightarrow&\delta^{\tau}(\omega-\omega^{\prime})\approx \frac{\tau}{2\pi}.
%\label{dt3}
%\eea

If $\vec{k}^{\prime}\neq\vec{k}$ and $\epsilon^{\prime}\neq\epsilon$, i.e. if the photon 
is scattered in a different direction than the incident one, Eq.~(\ref{ro3}) becomes
$$
\bra{\vec{k}^{\prime},\epsilon^{\prime}}U_I(t+\Delta t,t)\ket{\vec{k},\epsilon} \; =
\;\;\;\;\;\;\;\;\;\;\;\;\;\;\;\;\;\;\;\;\;\;\;\;\;\;\;\;\;
$$
\be
- \, 2\pi i\frac{g_{\vec{k},\epsilon}g^{*}_{\vec{k}^{\prime},\epsilon^{\prime}}}{\Delta-i\frac{\gamma}{2}}
\delta^{\tau}(\omega_{k^{\prime}}-\omega_k)\ket{a}\bra{a}.
\ee
Therefore, the non-normalized final state of the atom, after the photon is scattered and then detected, is
\be\label{fs2}
\ket{\Psi_{k^\prime}}= 
 -2\pi i\frac{g_{\vec{k},\epsilon}g^{*}_{\vec{k}^{\prime},\epsilon^{\prime}}}{\Delta-i\frac{\gamma}{2}}
\delta^{\tau}(\omega_{k^{\prime}}-\omega_k)\ket{a}\braket{a}{\Psi} \; .
\ee
From the above equation it is clear that, after the interaction with the photon, 
the atomic state is projected on  the stable state $\ket{a}$, 
the same atomic state that was populated before the interaction.
Of course, if initially the atom is in state $\ket{b}$, according to our assumptions 
it does not interact with the photon, and the latter is then transmitted unaffected. 

The probability that after the photo-detection the atomic  state is $\ket{\Psi_{k^\prime}}$, is given by
the square norm $\|\ket{\Psi_{k^\prime}}\|^2$. This coincides with the probability that a photon 
is emitted in state $\ket{\vec{k}^{\prime},\epsilon^{\prime}}$ during a time interval $\tau$. 
Denoting $I_{\vec{k}^{\prime},\epsilon^{\prime}}(\Psi)=\|\ket{\Psi_{k^\prime}}\|^2$,
we have
\bea\label{prob}
I_{\vec{k}^{\prime},\epsilon^{\prime}}(\Psi)&=&
2\pi\tau\frac{\omega_{k^{\prime}}\omega_k}{(2\varepsilon_0 L^3)^2}
\frac{|\vec{d}_{ac}\cdot\an{\epsilon}|^2|\vec{d}_{ac}\cdot\an{\epsilon}^{\prime}|^2}
{\Delta^2+\frac{\gamma^2}{4}}\nonumber\\
&&\times\delta(\omega_{k^{\prime}}-\omega_k)\|\ket{a}\braket{a}{\Psi}\|^2,
\eea
where we have written $g_{\vec{k},\epsilon}$ according to Eq.~(\ref{g}), and we have
used the approximation
$\lim_{\tau\rightarrow\infty}\left[\delta^{\tau}(\omega_{k^{\prime}}-\omega_k)\right]^2=
\tau \delta(\omega_{k^{\prime}}-\omega_k)/2\pi$.
Because we are assuming that only the direction of the scattered photon is revealed, we have 
to trace over the polarization components and to sum on the energy spectrum. 
Thus, the probability that the photon is emitted along direction $\an{k}^{\prime}$ is
\bea
I_{\an{k}^{\prime}}(\Psi)&=& 
\sum_{\epsilon^{\prime}} \frac{L^3}{(2\pi)^3}\int_{0}^{\infty}d k^{\prime}k^{\prime 2}
I_{\vec{k}^{\prime},\epsilon^{\prime}}(\Psi)\nonumber\\
&=&\sum_{\epsilon^{\prime}} \frac{L^3}{(2\pi)^3}\int_{0}^{\infty}d \omega^{\prime}\omega^{\prime 2}
I_{\vec{k}^{\prime},\epsilon^{\prime}}(\Psi)\;,\label{prob2}
\eea  
where $|\vec{k}|=k=\omega_k$.
The integral in the above equation is easily evaluated, yielding
\be
I_{\an{k}^{\prime}}(\Psi)=\sum_{\epsilon^{\prime}}
\frac{\omega_{k}^4 |\vec{d}_{ac}\cdot\an{\epsilon}|^2|\vec{d}_{ac}\cdot\an{\epsilon}^{\prime}|^2}
{(2\pi)^2 4\varepsilon_{0}^2 L^3 (\Delta^2+\frac{\gamma^2}{4})}\tau\|\ket{a}\braket{a}{\Psi}\|^2.
\ee 
In order to perform the sum over $\epsilon^{\prime}$, let us assume that the transition 
dipole moment is directed along the $z$ axis, and that the wave vector of the incident photon 
points in the 
positive $y$'s direction. Then we can write $\vec{d_{ac}}=d_{ac}\an{z}$, so that 
$\vec{d_{ac}}\cdot\an{\epsilon}=d_{ac}\epsilon_z=d_{ac}$, since $|\epsilon_z|=1$, 
and $\vec{d_{ac}}\cdot\an{\epsilon}^{\prime}=d_{ac}\epsilon^{\prime}_z$.
Moreover, making use of the standard relation \cite{cohen92}
$\sum_{\epsilon^{\prime}}|\vec{d}_{ac}\cdot\an{\epsilon}^{\prime}|^2
= d_{ac}^2-\vec{d}_{ac}\cdot\vec{k}^{\prime}/\omega_{k}^2
= d_{ac}^2(1-\an{k}^{\prime 2}_{z})$,
%\begin{eqnarray*}
%\sum_{\epsilon^{\prime}}|\vec{d}_{ac}\cdot\an{\epsilon}^{\prime}|^2
%&=& d_{ac}^2-\frac{\vec{d}_{ac}\cdot\vec{k}^{\prime}}{\omega_{k}^2}\\
%&=& d_{ac}^2(1-\an{k}^{\prime 2}_{z}),
%\end{eqnarray*}
and taking  $\an{k}^{\prime}\equiv(\sin\theta\cos\varphi,\sin\theta\sin\varphi,\cos\theta)$,
where $\theta$ and $\varphi$ are the azimuth and polar angle respectively,
we finally have
\be\label{prob3}
I_{\an{k}^{\prime}}(\Psi)=
\frac{\omega_{k}^4 d_{ac}^4(1-\cos^2\theta)}
{(2\pi)^2 4\varepsilon_{0}^2 L^3 (\Delta^2+\frac{\gamma^2}{4})}\tau\|\ket{a}\braket{a}{\Psi}\|^2.
\ee

It is now convenient for our purposes
to express $I_{\an{k}^{\prime}}(\Psi)$ in terms of the interaction strength and of the 
spontaneous emission rate. Actually, since $\epsilon_z=1$, from Eq.~(\ref{g}) we can write
$g_{\vec{k},\epsilon}=
-d_{ac}\sqrt{\omega_k/2\varepsilon_0 L^3}\equiv g$, 
and because $\gamma=\omega_{k}^{3}d_{ac}^2/3\pi\varepsilon_0$, Eq.~(\ref{prob3}) becomes
\be
I_{\an{k}^{\prime}}(\Psi)= 
\frac{3}{8\pi}\frac{\gamma |g|^2(1-\cos^2\theta)}{\Delta^2+\frac{\gamma^2}{4}}
\tau\|\ket{a}\braket{a}{\Psi}\|^2.
\ee  

We can now write the explicit form of the reset operator:
\be\label{ro4}
\an{R}_{\an{k}^\prime}=\sqrt{\xi^{*}_{\an{k}^{\prime}}}\ket{a}\bra{a}\;,
\ee
where $\xi^{*}_{\an{k}^{\prime}}=
3\gamma |g|^2(1-\cos^2\theta)\tau/8\pi(\Delta^2+\gamma^2/4)$.

It is immediate to generalize this form to the case of an ensembles of $N$ 
atoms. We simply have
\be\label{ro5}
\an{R}_{\an{k}^\prime}=
\sqrt{\xi^{*}_{\an{k}^{\prime}}}\sum_{n=1}^{N}\ket{a}_{n}\!\bra{a}\;,
\ee
and the probability that the photon is emitted along direction $\an{k}^{\prime}$ is
\be\label{Ik}
I_{\an{k}^{\prime}}(\Psi)=\|\an{R}_{\an{k}^\prime}\ket{\Psi}\|^2.
\ee
Thus, after the photon has interacted with the atom, whose initial state is denoted by $\ket{\Psi}$, 
and it has been detected, the updated atomic state, properly normalized, reads
\be
\ket{\Psi^{\prime}}=\frac{\an{R}_{\an{k}^\prime}\ket{\Psi}}{\sqrt{I_{\an{k}^{\prime}}(\Psi)}}.
\ee
\subsection*{The effective time evolution operator}\label{ETO}

Let us consider now the case in which the photon is emitted in the same state of 
the incident one, i.e., when $\vec{k}=\vec{k}^{\prime}$ and $\epsilon=\epsilon^{\prime}$. 
Since for $\omega\approx\omega^{\prime}$ we have
$\delta^{\tau}(\omega-\omega^{\prime})\approx \tau/2\pi$,
Eq.~(\ref{ro3}) 
can be rewritten as
\be\label{eteo1}
\bra{\vec{k},\epsilon}U_I(t+\Delta t,t)\ket{\vec{k},\epsilon}=
\mathbb{I}-i\frac{|g|^2}{\Delta-i\frac{\gamma}{2}}\tau\ket{a}\bra{a}\;.
\ee
If we take $\Delta\gg\gamma/2$, to first order in $\gamma/\Delta^2$, we have 
$1/(\Delta-i\gamma/2)\approx 1/\Delta -i\gamma/2\Delta^2$. 
Adhering to this approximation in Eq.(\ref{eteo1}), we have 
$1- i|g|^2\tau/(\Delta-i\gamma/2)\approx 1-i|g|^2\tau/\Delta
-|g|^2\gamma\tau/2\Delta^2
\approx e^{-i|g|^2\tau/\Delta-|g|^2\gamma\tau/2\Delta^2}$.
Hence, the action of the time evolution operator on the state of the photon-atom 
system is simply 
to multiply this state by an exponential factor, i.e., 
$\ket{\vec{k},\epsilon}\ket{a}\stackrel{\tau}{\longrightarrow}
e^{-i|g|^2\tau/\Delta-|g|^2\gamma\tau/2\Delta^2}\ket{\vec{k},\epsilon}\ket{a}$.
By extending this description to the general case of $N$ atoms,
we can write the following effective non-Hermitian Hamiltonian
\be\label{eff.hamiltonian}
H_{eff}=\sum_n \frac{|g|^2}{\Delta}\ad{a}_I\an{a}_I\ket{a}_n\!\bra{a}
-i\frac{|g|^2\gamma}{2\Delta^2}\ad{a}_I\an{a}_I\ket{a}_n\!\bra{a}, 
\ee
where we have used $\ket{\vec{k},\epsilon}\equiv\ad{a}_I\ket{0}$.  
The physical meaning of the above expression is quite clear. The first term in the right-hand side 
is the phase-shift, proportional to the total population of state $\ket{a}$,
caused by the off-resonant interaction between light and atoms. 
The second term 
represents the depletion of the field mode $\an{a}_I$ due to the spontaneous scattering.
Thus, in the case the photon is off-resonantly transmitted, 
the atomic state evolution can be simply written as 
$\ket{\Psi^{\prime}}\propto e^{(-i|g|^2\tau/\Delta-|g|^2\gamma\tau/2\Delta^2)\sum_n\ket{a}_n\!\bra{a}}\ket{\Psi}$. 
 
We can introduce a simplified description that elucidates 
the role of the probabilities for the alternative evolutions of the atomic state.
Let us consider only one atom, whose initial state is $\ket{a}$ and, hence, will surely interact
with the photon. If the photon is scattered 
in the same mode of the incident one, after an interval $\tau$, the atomic state is
$e^{-i|g|^2\tau/\Delta-|g|^2\gamma\tau/2\Delta^2}\ket{a}$. 
We can write the damping term as 
$e^{-|g|^2\gamma\tau/2\Delta^2}\approx 1-|g|^2\gamma\tau/2\Delta^2\approx
\sqrt{1-|g|^2\gamma\tau/\Delta^2}$. Hence, we have (in the case of one atom)
\be\label{inter}
\ket{\Psi^{\prime}}
=\frac{\sqrt{1-\frac{|g|^2\gamma}{\Delta^2}\tau}e^{-i\frac{|g|^2}{\Delta}\tau}\ket{a}}
{\|\cdot\|}\;,
\ee
where $\|\cdot\|$ is the appropriate normalization factor.
The factor in square in the above equation can be consistently interpreted 
as the probability amplitude that no spontaneous scattering occurs.
In fact, to first order in $\gamma/\Delta^2$, 
the probability that the photon is scattered in the $\an{k}^{\prime}$ direction by an atom in state
$\ket{a}$ is:
%\be\label{xi}
$\xi_{\an{k}^{\prime}}\equiv\|I_{\an{k}^{\prime}}(a)\|=3|g|^2\gamma\tau(1-\cos^2\theta)/8\pi\Delta^2$.
%\ee 
By integrating over the whole solid angle, we recover the probability 
to have spontaneous scattering: 
\bea\label{prob4}
\xi_s&=&\int\|I_{\an{k}^{\prime}}(a)\|d\Omega \nonumber\\
&=& \int_{0}^{2\pi}d\varphi\int_{0}^{\pi}
\frac{3}{8\pi}\frac{|g|^2\gamma\tau(1-\cos^2\theta)}{\Delta^2}\sin\theta d\theta\nonumber\\
&=&\frac{|g|^2\gamma}{\Delta^2}\tau\label{scprob}\;.
\eea
Consequently, the above interpretation of the factor in square root 
in the right-hand side of Eq.~(\ref{inter}) follows.
  
This result can be generalized by induction to the case of $N$ atoms. 
If $n_{a}$ is the number of atoms populating state $\ket{a}$, 
using Eqs.~(\ref{ro5},\ref{Ik})
the probability of spontaneous scattering can be written as
$\xi_N=\xi_{s}n_{a}^2$. Therefore the square root term becomes 
$\sqrt{1-\xi_s n_{a}^{2}}$. Obviously, this approximation holds 
as long as $\xi_s n_{a}^{2}\ll 1$.

\section{Creation of entanglement by photo-detection }\label{CEP}

Let us introduce the atomic spin operators for the stable states:
$j_{nz}=(\ket{b}_{n}\!\bra{b} - \ket{a}_{n}\!\bra{a})/2$,
$j_{n+}=\ket{b}_{n}\!\bra{a}/2$,
$j_{n-}=\ket{a}_{n}\!\bra{b}/2$.
%\begin{eqnarray}
%j_{nz} & = &
%\frac{1}{2}\left(\ket{b}_{n}\!\bra{b} - \ket{a}_{n}\!\bra{a}\right) \\
%j_{n+} & = & \frac{1}{2}\ket{b}_{n}\!\bra{a}\\
%j_{n-} & = & \frac{1}{2}\ket{a}_{n}\!\bra{b},
%\end{eqnarray} 
The reset operator (\ref{ro5}) and the effective Hamiltonian (\ref{eff.hamiltonian}) become
\bea
\an{R}_{\an{k}^\prime}&=&
\sqrt{\xi_{\an{k}^{\prime}}}\sum_{n=1}^{N}\left(\frac{1}{2} -j_{nz}\right)\label{ro6},\\
H_{eff}&=&\frac{|g|^2}{\Delta}\sum_n \left(\frac{1}{2} -j_{nz}\right)a_{I}^{\dag}a_{I}+\nonumber\\
&&
-i\frac{|g|^2\gamma}{2\Delta^2}\sum_n \left(\frac{1}{2} -j_{nz}\right)a_{I}^{\dag}a_{I}\label{eff.hamiltonian2}\;.
\eea

Since we are concerned with ensembles of atoms in which each atom is 
initially prepared in the same state and interacts in an identical way
%the interaction 
with the surrounding environment, 
%is identically the same for all atoms, 
the collective atomic state
preserves the full permutation symmetry. It is therefore convenient to expand 
this collective state in the eigenstates of the effective collective angular momentum:
$\ket{\Psi}= \sum_{M=-J}^{J}\mathcal{A}_{M}\ket{J,M}$,
where $J=N/2$ is the total angular momentum, and 
$M = \left(n_{b}-n_{a}\right)/2$ is the eigenvalue of
the operator $J_{z}= \sum_{n}^{N} j_{nz}$. Collective raising and
lowering operators, and the corresponding Cartesian $x$- and $y$-
components of the collective angular momentum, are defined 
as similar sums over all atoms in the sample. 
Notice that if the atoms are prepared to populate only the two stable states $\ket{a}$ and $\ket{b}$, 
the total population of state $\ket{a}$ can be written as:   
$\sum_{n=1}^{N}\left(1/2  -M\right)=\left(N/2 - M\right)\nonumber
= (n_{a} + n_{b})/2 - (n_b - n_a)/2 = n_a$.  
%\bea
%\sum_{n=1}^{N}\left(\frac{1}{2} -M\right)&=&\left(\frac{N}{2} - M\right)\nonumber\\
%&=& \left(\frac{n_{a} +  
%n_{b}}{2} - \frac{n_b - n_a}{2}\right) = n_a \label{na} 
%\eea.

%\subsection{Creation of entanglement by photo-detection }\label{CEP}

The generalization of the atomic spin formalism to the case of two atomic samples 
with $N_1$ and $N_2$ atoms respectively, is straightforward. 
The initial disentangled state of the atomic samples
is expanded in product-state wave functions of the two ensembles
$\ket{\Psi} = \sum_{M_{1},M_{2}}\mathcal{A}_{M_1,M_2}\ket{M_{1},M_{2}}$,
%\begin{equation}\label{is}
%\ket{\Psi} = 
%\sum_{M_{1},M_{2}}\mathcal{A}_{M_1,M_2}\ket{M_{1},M_{2}},
%\end{equation}
where $|M_1,M_2\rangle \equiv 
|J_1=N_1/2,M_1\rangle\otimes|J_2=N_2/2,M_2\rangle$, and 
we assume an initial product state $\mathcal{A}_{M_1,M_2} =
\mathcal{A}_{M_1}\mathcal{A}_{M_2}$ before we let 
the system interact with the incident field.
The total reset operator is simply the sum of the reset operators 
acting on a single ensemble of atoms as expressed by Eq.~(\ref{ro6}),
Hence, we have 
$$
\an{R}_{\an{k}^\prime}\ket{M_1,M_2}=
\;\;\;\;\;\;\;\;\;\;\;\;\;\;\;\;\;\;\;\;\;\;\;\;\;\;\;\;\;\;\;\;\;\;\;\;\;\;\;\;\;\;\;\;\;\;\;\;
$$
\be
\sqrt{\xi_{\an{k}^\prime}}
\left[\frac{N_1+N_2}{2}-(M_1+M_2)\right]\ket{M_1,M_2}\label{ro7},
\ee
where $(N_1+N_2)/2 -(M_1+M_2)=n_a\equiv\mathcal{N}_{(M_1+M_2)}$ 
is the total population of the atomic state $\ket{a}$.  

After the photon has gone through the interferometric set-up
and has been detected, the atomic state will be modified.
In particular, if the state of the photon entering the interferometer is
$\ket{+}_{ph} = (\ket{F}_{ph}+\ket{I}_{ph})/\sqrt{2}$,
where  $\ket{F}_{ph}$ and $\ket{I}_{ph}$ denote the states of the photons that 
follow respectively the upper or the lower path of the interferometer and
$\ket{\pm}_{ph} = (\ket{F}_{ph}\pm\ket{I}_{ph})/\sqrt{2}$ 
are the photon states revealed by the detectors  $D_{+}$ and $D_{-}$,
the initial state of the system photon+atoms is 
$\ket{\phi}_{ph+at}= \ket{\Psi} \otimes (\ket{F}_{ph}+\ket{I}_{ph})/\sqrt{2}$.
The analysis presented in the previous sections provides us with the recipe to work out
the evolution of the \emph{interacting} component of $\ket{\phi}_{ph+at}$.
Therefore, by using Eq.~(\ref{ro7}) and the multi-atom generalization of Eq.~(\ref{inter}),
after the photon has interacted, and before it is detected,
we can formally write the photon+atoms state vector as
$$
\ket{\phi}_{ph+at}^{\prime}=
\sum_{M_{1},M_{2}}\left[\frac{\mathcal{A}_{M_1,M_2}}{\sqrt{2}}\ket{F}_{ph}\ket{M_1,M_2}
\right.
$$
\bea
&&+\frac{\mathcal{A}_{M_1,M_2}}{\sqrt{2}}
\sqrt{1-\xi_{s}\mathcal{N}_{(M_1+M_2)}^2}\nonumber\\
&&\times e^{-i\frac{|g|^2}{\Delta}\tau\mathcal{N}_{(M_1+M_2)}}\ket{I}_{ph}\ket{M_1,M_2}
\nonumber\\
&&+\left.\frac{\mathcal{A}_{M_1,M_2}}{\sqrt{2}}
\sqrt{\xi_{\an{k}^\prime}}\mathcal{N}_{(M_1+M_2)}\ket{S}_{ph}\ket{M_1,M_2}
\right]\nonumber\\
&=& \sum_{M_{1},M_{2}}
\left[\frac{1+\sqrt{1-\xi_{s}\mathcal{N}_{(M_1+M_2)}^2}
e^{-i\frac{|g|^2}{\Delta}\tau\mathcal{N}_{(M_1+M_2)}}}{2}\right.\nonumber\\
&&\times\mathcal{A}_{M_1,M_2}\ket{+}\ket{M_1,M_2} \nonumber\\
&&+\frac{1-\sqrt{1-\xi_{s}\mathcal{N}_{(M_1+M_2)}^2}
e^{-i\frac{|g|^2}{\Delta}\tau\mathcal{N}_{(M_1+M_2)}}}{2}\nonumber\\
&&\times\mathcal{A}_{M_1,M_2}\ket{-}\ket{M_1,M_2}
\nonumber\\
&&\left.+\frac{\mathcal{A}_{M_1,M_2}}{\sqrt{2}}
\sqrt{\xi_{\an{k}^\prime}}\mathcal{N}_{(M_1+M_2)}\ket{S}_{ph}\ket{M_1,M_2}\right]\nonumber\;,
\eea
where $\ket{S}_{ph}$ is the state of the photon scattered in the direction $\an{k}^{\prime}$.
Finally, to obtain the properly normalized state, we must sum over every possible 
spontaneously-scattered photon state.

From the above equation is clear how the atomic state amplitudes change after the photo-detection.
We must consider three cases:
\begin{itemize}
 \item If the photon is detected by $D_+$ or $D_-$, then the updating procedures for the atomic 
       state amplitudes are respectively
       $$
       \mathcal{A}_{M_1,M_2}\longrightarrow\frac{\mathcal{A}_{M_1,M_2}}{2}
       \;\;\;\; \;\;\;\; \;\;\;\; \;\;\;\;\;\;\;\;\;\;\;\;\;\;\;\;\;\;\;\;\;\;\;\;\;\;\;\;
       $$
       \be
       \times\left[1+\sqrt{1-\xi_{s}\mathcal{N}_{(M_1+M_2)}^2}\;
       e^{-i\frac{|g|^2}{\Delta}\tau\mathcal{N}_{(M_1+M_2)}}\right]\;,
       \;\;\;\;\;\;\;\;\label{upa1}
       \ee
       $$
       \mathcal{A}_{M_1,M_2}\longrightarrow \frac{\mathcal{A}_{M_1,M_2}}{2}
       \;\;\;\; \;\;\;\; \;\;\;\; \;\;\;\;\;\;\;\;\;\;\;\;\;\;\;\;\;\;\;\;\;\;\;\;\;\;\;\;
       $$
       \be 
       \times\left[1-\sqrt{1-\xi_{s}\mathcal{N}_{(M_1+M_2)}^2}\;
       e^{-i\frac{|g|^2}{\Delta}\tau\mathcal{N}_{(M_1+M_2)}}\right]\;,
       \;\;\;\;\;\;\;\;\label{upa2}
       \ee 
       and we get the non-normalized states
       $$
       \ket{\Psi_{\pm}}= \sum_{M_{1},M_{2}}
       \frac{\mathcal{A}_{M_1,M_2}}{2}\ket{M_1,M_2}
       \;\;\;\;\;\;\;\;\;\;\;\;\;\;\;\;\;\;\;\; 
       $$
       \be\label{ups1}
       \times\left[1\pm\sqrt{1-\xi_{s}\mathcal{N}_{(M_1+M_2)}^2}\;
       e^{-i\frac{|g|^2}{\Delta}\tau\mathcal{N}_{(M_1+M_2)}}\right].
       \ee 
       The probabilities that the photon is detected in $D_+$ or $D_-$ 
       are simply $P_{\pm}=\|\ket{\Psi_{\pm}}\|^2$, 
       and the normalized atomic state is $\ket{\Psi_{\pm}^{\prime}}=\ket{\Psi_{\pm}}/\|\ket{\Psi_{\pm}}\|$.
 \item If the photon is spontaneously scattered, the atomic state amplitudes become
       \be\label{upa3}
       \mathcal{A}_{M_1,M_2}\longrightarrow
       \frac{\mathcal{A}_{M_1,M_2}}{\sqrt{2}}\sqrt{\xi_{\an{k}^\prime}}\mathcal{N}_{(M_1+M_2)}\;,
       \ee
       and the corresponding non-normalized state is
       \be\label{ups2}
       \ket{\Psi_{S}}=\sum_{M_{1},M_{2}}
       \frac{\mathcal{A}_{M_1,M_2}}{\sqrt{2}}\sqrt{\xi_{\an{k}^\prime}}\mathcal{N}_{(M_1+M_2)}\ket{M_1,M_2}.
       \ee
       In this case, the probability that the photon is scattered in the $\an{k}^{\prime}$ direction 
       is  $P_{\an{k}^{\prime}}=\|\ket{\Psi_S}\|^2$, and the normalized state reads 
       $\ket{\Psi_{S}^{\prime}}=\ket{\Psi_{S}}/\|\ket{\Psi_{S}}\|$. 
       Let us notice that the probability that the photon is scattered in any direction is simply
       $P_{s}=\int\|\ket{\Psi_S}\|^2 d\Omega$, and then $P_{+}+P_{-}+P_{s}=1$.      
\end{itemize}   

Form Eqs.~(\ref{ups1}) and (\ref{ups2}), the entangled nature of the final atomic states is evident:
because the amplitudes ``evolved'' into functions of the sum $(M_1+M_2)$,
it is not possible to express the atomic state vector as a direct product of single atomic sample states.
Every time a photon is detected, the updating procedure will be repeated  
according the following \emph{simulation scheme}:
\begin{enumerate}
 \item The probability $P_{s}$ is determined and compared with 
       a uniformly distributed random number (UDRN) $r_1$.
 \item If $P_{s}<r_1$, then we calculate the conditional 
       probability that a photon is detected in $D_+$, $P_{+}^{c}=P_{+}/(1-P_{s})$,
       and compare its value with an other UDRN  $r_2$:
       \begin{description}
        \item{i)} If $P_{+}^{c}>r_2$, then the atomic state amplitudes change according to Eq.~(\ref{upa1}); 
        \item{ii)} If $P_{+}^{c}<r_2$, then the atomic state amplitudes change according to Eq.~(\ref{upa2}). 
       \end{description}
 \item  If $P_{s}>r_1$, we identify in which direction the photon is scattered 
        by determining the azimuthal angle $\theta$, whose density distribution function is given by 
        $2\pi\xi_{\an{k}^\prime}/\xi_{s}$, 
        and the atomic state amplitudes are updated using  Eq.~(\ref{upa3}).
 \item  Finally, we properly normalize the resulting atomic state vector and the process restarts.       
\end{enumerate}

After $N_{ph}$ photons have been detected, if we denote with $N_S$ 
the number of spontaneously scattered photons revealed by the surrounding detectors, 
and if $N_+$ and $N_-$ are the photons detected by $D_+$ and $D_-$ respectively, 
such that $N_S+N_++N_-=N_{ph}$, the atomic state vector will be
$$
\ket{\Psi^{\prime}}=\frac{1}{\|\cdot\|}\sum_{M_{1},M_{2}}\mathcal{A}_{M_1,M_2}
\;\;\;\;\;\;\;\;\;\;\;\;\;\;\;\;
$$
\bea
&&\times\left(\prod_{n_s=1}^{N_S}\sqrt{\frac{\xi_{\an{k}^\prime}^{n_s}}{2}}\right)\mathcal{N}^{N_S}_{(M_1+M_2)}
\nonumber\\
&&\times\mathcal{F}^{N_+}_{+}(M_1+M_2)\mathcal{F}_{-}^{N_-}(M_1+M_2)\ket{M_1,M_2},\nonumber
\eea
where 
$$
\mathcal{F}_{\pm}(M_1+M_2)=\frac{1 \pm \sqrt{1-\xi_{s}\mathcal{N}_{(M_1+M_2)}^2}
e^{-i\frac{|g|^2}{\Delta}\tau\mathcal{N}_{(M_1+M_2)}}}{2} ,
$$
and $\xi_{\an{k}^\prime}^{n_s}$ is the scattering factor corresponding to the $n_s$-th detection 
along the different directions $\an{k}^{\prime}$'s.

\section{Numerical Simulations}\label{NS}

In this section we will implement in a numerical simulation the detection model described in the 
preceding section.
Following the measurement scheme introduced in Ref.~\cite{noi3},
we will consider two atomic samples having the same number $N$ of atoms and whose initial state is the 
eigenstate of the spin operator $J_{x1}+J_{x2}$, with eigenvalue  $J_1+J_2 = 2J =N$. 
This means that all the atoms are initially prepared in the superposition state  
$(\ket{a}+\ket{b})/\sqrt{2}$. The corresponding amplitudes in the basis eigenstates of the 
$z$-component of the collective angular momentum operators
$|J,M_i\rangle$ are
\be\label{iamp}
\mathcal{A}_{M_i} =
\left(\frac{1}{2}\right)^{J}\sqrt{\frac{(2J)!}{(J+M_{i})!(J-M_{i})!}}
\;\;,
\ee
where $i=1,2$. Expressed in terms of the number of atoms in state
$\ket{a}$ in each sample, $n_{a}^{i}$, the square of the 
amplitude (\ref{iamp}) is
$|\mathcal{A}_{M_i}|^2 = (1/2)^{N}N!/(N-n_{a}^{i})!(n_{a}^{i})!$. This means that
in each ensemble the atoms are distributed in state $\ket{a}$ and
$\ket{b}$ according to a binomial distribution with probability
$1/2$.

In what follows, we will present the results of two kinds of numerical simulation where
the entanglement of the two atomic samples is monitored. 
The first scheme corresponds to consecutive measurements of the angular momentum operators 
$J_{z1}+J_{z2}$ and $J_{y1}-J_{y2}$, while the second corresponds to a simulated
evolution in which
continuous opposite rotations are applied to the atomic spin of the two samples as the 
photo-detection proceeds. Thereby we are considering a continuous exchange of the operators 
$J_{z1}+J_{z2}$ and $J_{y1}-J_{y2}$.
We will compare the case in which the spontaneous scattering is neglected, \emph{i.e.} $\xi_s=0$, 
with that in which the spontaneous emission rate $\gamma$ 
is much smaller, but not negligible, than the detuning $\Delta$.
Specifically, we will take $\Delta=150\gamma$.

\subsection*{ Consecutive measurements of
$J_{z1}+J_{z2}$ and $J_{y1}-J_{y2}$}\label{numerical}

Starting from an initial state whose amplitude is given by the product of those of each samples
expressed in Eq.~(\ref{iamp}), we consider 
a series of photo-detections that project the state according 
to Eqs.~(\ref{ups1}) and (\ref{ups2}).
     
As the photo-detection proceeds, the uncertainty in $M_{12}=M_1+M_2$, 
denoted with $\Delta^2(J_{1z}+J_{2z})$,
goes to zero \cite{noi3}, i.e., after a large number of photons has been detected 
the state of the two ensembles approximates an eigenstate of $J_{z1}+J_{z2}$. 
This fact is true also if spontaneous scattering is taken in account. 
Namely, we have seen that the detection of scattered photons measures the atomic state 
population given by $(N_1+N_2)/2 - (M_1+M_2)$, just as the phase shift measurement 
carried out by detectors $D_+$ and $D_-$ does.
If we take $\delta=g^{2}/\Delta$, we assume 
a phase angle $\delta\tau=0.1$, corresponding to a large (and hardly experimentally
available) phase shift on the atomic state due 
to interaction with a freely propagating field. However, in our simulations
a smaller and more realistic value of $\delta\tau$ just implies 
that more photons have to be detected to achieve the same 
reduction in the variance $\Delta^2(J_{1z}+J_{2z})$.

To quantify the pure-state entanglement between the two samples 
we will use the \emph{entropy of entanglement}, 
defined  by Bennett et al. \cite{bennett96} to measure the entanglement
of two systems in a pure state $|\Psi\rangle$:
\be
\mathcal{E} = -Tr(\rho_1 \log_2\rho_1)=-Tr(\rho_2 \log_2\rho_2),
\ee
where $\rho_{1}=Tr_{2}\ket{\Psi}\bra{\Psi}$ is the reduced density
matrix of system 1, and analogously for $\rho_2$. If $N$ is the number of
atoms in each sample, and we restrict ourselves to states which are
symmetric under permutations inside the samples, the quantity
$\mathcal{E}$ takes values between zero for a product state, and
$\log_2(N+1)$ for a maximally entangled state of the two samples.

\begin{figure}[hbtp]
\begin{center}
\includegraphics[width=7.5cm]{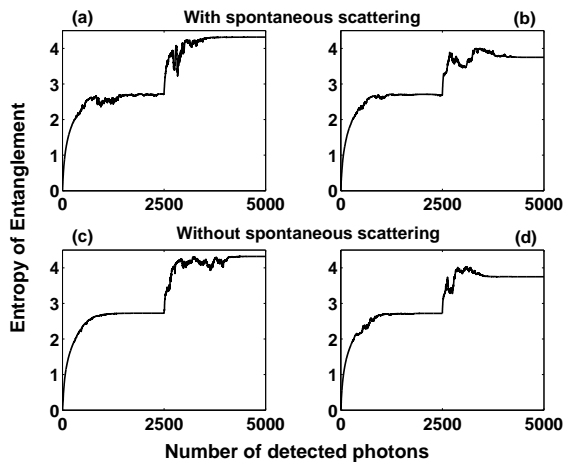}
\end{center} 
%%\vspace{2cm}
\caption{\small{Entanglement of two samples each with $N=20$ atoms. 
After detection of the first
$N_{ph}=2500$ photons, the samples are rotated $\pm 90$ degrees in spin space,
and the subsequent detection events lead, typically, 
to further entanglement. Figs.~.(a) and (b) show results of two different
simulation records for $\Delta=150\gamma$.  
Figs.~(c) and (d) show results of two different simulation records 
for $\gamma=0$. In both cases, $\delta\tau = 0.1$.}}
\label{ex1}
\end{figure}

Fig.~\ref{ex1} shows the evolution of entanglement between the two atomic samples 
obtained in four numerical simulations of the consecutive measurement of  
$J_{z1}+J_{z2}$ and $J_{y1}-J_{y2}$. For a number of detected photons $N_{ph}\leq 2500$, 
the state vector evolves according to the analysis exposed above, i.e., it goes approximately 
to an eigenstate of $J_{z1}+J_{z2}$. We then apply opposite rotations to the atomic samples
and proceed with similar measurements as before, which effectively measure $J_{y1}-J_{y2}$.    
After the rotations, the full permutation symmetry of the system is broken, 
i.e., the total angular momentum $J^2 = J_{x}^2+J_{y}^2+J_{z}^2$ is not
conserved. The final state of the two samples is still a simultaneous
eigenstate of both $J_{1}^2$ and $J_{2}^2$, but with
different values of the total angular momentum.
 
Figs.~\ref{ex1}-(a) and -(b) show the results of two simulations with
spontaneous scattering. Fig.~\ref{ex1}-(a) represents an example
of the evolution of the entropy in which the final value ($\mathcal{E}_{N_{ph}}= 4.3165$) 
is very close to that of the maximally entangled state
($\log_{2}(20+1)=4.3923$). In Fig.~\ref{ex1}-(b), we show a typical evolution, where
the final value of the entanglement is well below the maximum ($\mathcal{E}_{N_{ph}}= 3.7494 $).
Figs.~\ref{ex1}-(c) and -(d) show the corresponding evolution of the 
entanglement when spontaneous scattering is neglected. In Fig.~\ref{ex1}-(c), the final value of the entropy
is $\mathcal{E}_{N_{ph}}= 4.3194$, while that reached in the simulation represented in Fig.~\ref{ex1}-(d) 
is $\mathcal{E}_{N_{ph}}= 3.75$.

\begin{figure}[hbtp]
\begin{center}
\includegraphics[width=7.5cm]{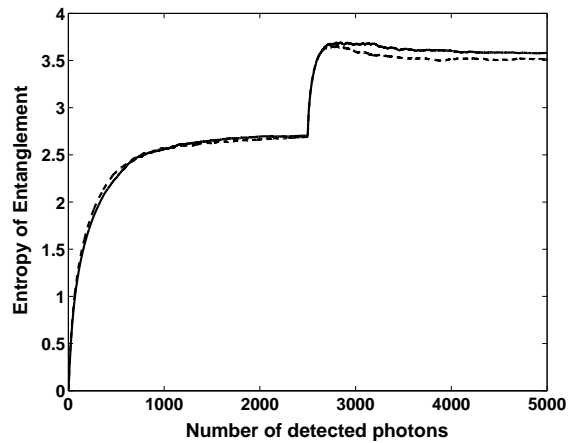}
\end{center} 
%%\vspace{2cm}
\caption{\small{Average of the entanglement over 100 simulations of consecutive 
measurement of  $J_{z1}+J_{z2}$ and $J_{y1}-J_{y2}$.
Dashed line: average evolution when $\Delta=150\gamma$.
Full line:  average evolution when $\gamma=0$. Here
$\delta\tau=0.1$ as in Fig.~\ref{ex1}.}}
\label{ex2}
\end{figure}

The behavior of the two types of simulated evolutions are very similar, as confirmed in Fig.~\ref{ex2},
where the average over 100 simulations for each type of interaction scheme is presented. Actually, 
in the first part of the simulation, for $N_{ph}<2500$, the two average evolutions are 
practically the same, reaching the same average value of the entanglement for $N_{ph}=2500$
($\mathcal{E}_{N_{ph}}= 2.7037$). 
For  $N_{ph}>2500$, after the rotations, despite the fact that the
difference of the final average entropies 
($\mathcal{E}_{N_{ph}}|_{\gamma=0}-\mathcal{E}_{N_{ph}}|_{\gamma\neq 0}=0.064$)
is smaller than the statistical fluctuations 
($\Delta\mathcal{E}_{N_{ph}}|_{\gamma=0}=0.47$), 
the average value of the entanglement in  
simulations with spontaneous scattering are constantly below those in which the 
spontaneous scattering is neglected. We will see that this feature 
is more evident in numerical simulations of measurements with continuous rotation 
of spin components.

In the majority of the simulations, for both interaction schemes, the second round of measurements 
increases the entanglement.
This fact can be qualitatively explained by noting that the approximate eigenstate
of $J_{z1}+J_{z2}$, produced by the detection, has amplitudes on different 
$M_1$ and $M_2$ states with $M_1+M_2$ fixed by the measured value, 
but the eigenspace is degenerate, and the amplitudes are
simply proportional to the ones in the initial state.
Due to the initial binomial distribution on $M_1$ and $M_2$,
the distribution over, e.g., $M_1$ will therefore have a width
of approximately $\sqrt{N}$. The reduced density matrix has the 
corresponding number of non-vanishing
populations, suggestive of $\mathcal{E} \sim \log_{2}\sqrt{N}
=0.5 \log_{2}N$, half of the maximal value. This argument
accounts for the first plateau reached in Fig.~\ref{ex1} 

The measurement of the $z$-components causes a broadening of the distribution
of the eigenstates of $J_y$, compared to the initial distribution, which was
also binomial in that basis. The subsequent measurement
of $J_{y1}-J_{y2}$ will produce a state with $M_{1y}-M_{2y}$
fixed by the measurement, but within the degenerate space of 
states with this fixed value the distribution on $M_{1y}$ of
the reduced density matrix is broader than $\sqrt{N}$, and the 
entanglement is correspondingly larger. 

\subsection*{Measurements with continuous rotation of spin components}\label{MCRSP}

The above arguments lead naturally to consider continuous exchanges of 
the spin operator, in order to generate states with higher entanglement. In this subsection,
we present the results of simulations in which continuous opposite rotations 
are applied to the atomic spins of the two samples as the photo-detections proceed.
If $\alpha$ is the variable rotation angle, what is effectively measured is the observable
$\an{J}_{\alpha}= \cos{\alpha}(J_{z1}+J_{z2})+\sin{\alpha}(J_{y1}-J_{y2})$.
If we denote with $\mathcal{A}_{M_1M_2}^{N_{ph}}$ the wave function amplitude of the
system after $N_{ph}$ photons have been detected, the updated state
vector rotated  by the angle $\pm \alpha$ prior to the subsequent detection is
\bea
\ket{\Psi}_{N_{ph}}&=&
\frac{1}{\sqrt{C}}\sum_{M^{'}_{1},M^{'}_{2}}
\sum_{M_{1},M{2}}\mathcal{D}^{J_1}_{M_1M^{'}_1}\left(\alpha\right)
\nonumber\\
&&\times\mathcal{D}^{J_2}_{M_2M^{'}_2}\left(-\alpha\right)\mathcal{A}_{M_1M_2}^{N_{ph}}
\ket{M^{'}_{1},M^{'}_{2}},\nonumber
\eea   
where  
$D^{J_i}_{M_iM^{'}_i}\left(\alpha\right)=
\bra{M_i,J_i}e^{-iJ_{ix}(\alpha)}\ket{J_i,M_{i}^{'}}$
is the rotation matrix element for sample $i$ ($i=1,2$). 
The wave function
updating algorithm has the same structure as the one described in section
(\ref{CEP}).  
In a realistic experimental situation, one would detect photons according to
a Poisson process. With a constant rotation frequency induced,
e.g., by applying opposite DC magnetic fields onto the atoms,
this would lead to small rotation angles with an exponential
distribution law. For simplicity, however, we apply the same
small rotation angle $\alpha$ between each detection event.
      
\begin{figure}[hbt]
\begin{center}
\includegraphics[width=7.5cm]{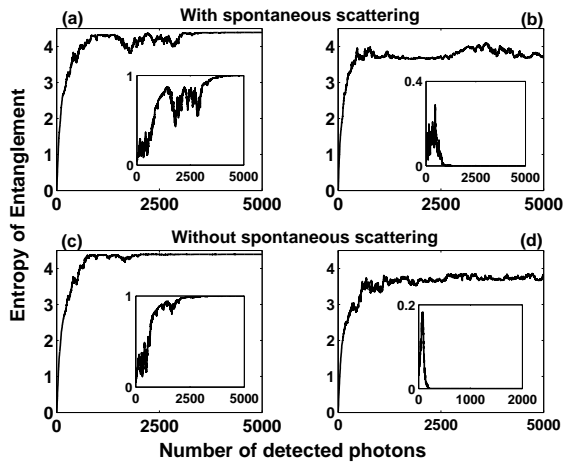}
\end{center} 
%%\vspace{2cm}
\caption{\small{Entanglement of atomic samples with 20 atoms each. 
The spins are rotated in opposite directions around the $x$-axis
by the angle $\alpha=\pi/5$ after each detection event.
Insets (a-d) show the evolution of the overlap
$|\bra{\Psi_0}\Psi\rangle_{N_{ph}}|^2$ between the state of the 
samples and the maximally entangled state. Figs.~(a) and (b) present
results of two different simulations in the case $\Delta=150\gamma$. 
Figs. (c) and (d) show analogue results for simulations in which
$\gamma=0$ (no spontaneous scattering).
In (a) and (c) $\ket{\Psi}_{N_{ph}}$ converges
toward $\ket{\Psi_0}$ and further detections have no effect on the
state vector. In (b) and (d) the state gradually loses its component
along $\ket{\Psi_0}$, and it subsequently evolves in the orthogonal 
subspace of $\ket{\Psi_0}$. Here, as in the previous simulations,
$\delta\tau=0.1$.}}
\label{ex3}
\end{figure}      
 
In Fig.~\ref{ex3} we show four numerical simulations of the entanglement evolution, two
for each interaction schemes. The rotation angle between subsequent photo-detection
events is $\alpha=\pi/5$.  
In the upper part of the panel, we provide results of two
numerical simulations with spontaneous scattering. Fig.~\ref{ex3}-(a) shows
the evolution in which the maximal value of the entanglement is reached, as is confirmed by
the inset of the figure showing the evolution of the overlap between the state vector of the two
atomic samples and the maximally entangled state
\be
|\Psi_0 \rangle =\frac 1{\sqrt{2J+1}}\sum_{M=-J}^J|M,-M\rangle.
\ee
In Fig.~\ref{ex3}-(b) we have a typical case in which the final state is orthogonal to
$|\Psi_0 \rangle$. The lower figures has been obtained
from simulations in which the spontaneous scattering has been neglected. In Fig.~\ref{ex3}-(c)
the evolution leading to the maximally entangled state is shown. A typical result is shown in 
Fig.~\ref{ex3}-(d).

Note that $|\Psi_0 \rangle$ is the only  joint eigenstate of the
operators $J_{1x}-J_{2x}$, $J_{1y}-J_{2y}$ and $J_{1z}+J_{2z}$ with null
eigenvalue \cite{noi3,BerrySanders1}. 
When we simulate the detection of phase shifts proportional to
$J_{1z}+J_{2z}$ and $J_{1y}-J_{2y}$, or combinations of these
operators, there is a nonvanishing probability that the state vector 
is gradually projected onto $|\Psi_0 \rangle$, and this state
is unaffected by all further photo-detection events, see Figs.~\ref{ex3}-(a)~and~-(c). 

If the state of the atomic samples has not collapsed into $|\Psi_0 \rangle$ after a large number of
photons have been detected, it instead gradually becomes orthogonal
to that state, and the maximally entangled
state is never attained, see the insets in Figs.~\ref{ex3}-(b)~and~-(d).  

\begin{figure}[hbt]
\begin{center}
\includegraphics[width=7.5cm]{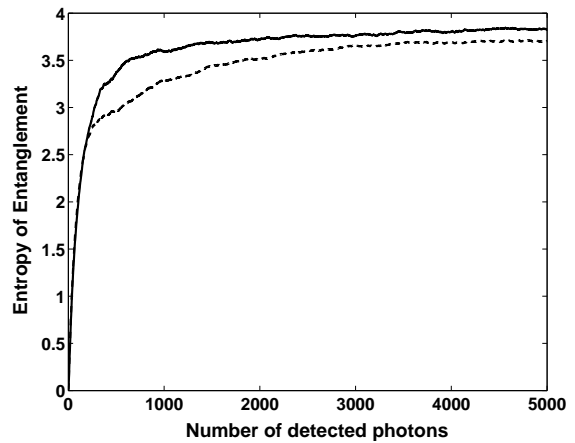}
\end{center} 
%%\vspace{2cm}
\caption{\small{Average of the Entanglement over 100 simulations of  
measurements with continuous rotations of the spin components.
Dashed line: average evolution when $\Delta=150\gamma$.
Full line: average evolution when $\gamma=0$. Here $\delta\tau=0.1$,
as in the previous figures.}}
\label{ex4}
\end{figure}     

In these simulations the qualitative behavior of the two types of interaction schemes
is again similar. As expected, there is a faster increase of the entanglement
and a more efficient evolution toward a maximally entangled state, compared with
the simulations of consecutive measurements of $J_{1z}+J_{2z}$ and $J_{1y}-J_{2y}$.
When spontaneous scattering is considered,
the evolution toward a stable regime appears to be slower, as is evident in
Fig.~\ref{ex4} where an average over 100 simulations is shown. Moreover, in this
case it appears clearly that the average values of the entanglement for evolutions 
with spontaneous scattering is constantly smaller than that obtained from evolutions
without spontaneous scattering. Even though this effect is
very weak, it seems to become evident when the rotational symmetry of the two systems is broken,
as seen in the previous section,
and it is suggestive of less correlated atomic samples due to photo-detections of spontaneous scattered photons.
Namely, when spontaneous scattering is considered, we have a wide spectrum of possible states in which the system
can be found. Hence, in detecting all the scattered photons, we gain information from a more disordered system
in comparison with the case in which spontaneous scattering is neglected. In the latter case 
there are only two possible states \cite{noi3} and less ignorance 
(the entropy is smaller) about the system.
Yet, the good news is that the effect of spontaneous scattering is not dramatically
destructive: our analysis provides indication that a QND interferometric scheme
to effectively entangle atomic samples should work very well also when this
unavoidable physical process is taken into account.    
      
It is worth noting that even when the final state does not coincide with  $|\Psi_0 \rangle$,
we observe an entanglement that is almost constant as the photo-detection goes on. 
This suggests the presence of families of
states, orthogonal to $|\Psi_0 \rangle$, with relatively
well-defined values of the operators $J_{1z}+J_{2z}$ and $J_{1y}-J_{2y}$, which is
allowed by Heisenberg's uncertainty relation as long as the expectation value of 
the operator $J_{1x}-J_{2x}$ is small \cite{noi3}. This interesting threshold behavior 
is unexpected, and at present not completely understood. 

\section{Conclusions and outlook} \label{Conclusions}

In this paper we have analyzed the effects of the spontaneous scattering on the entanglement evolution 
of two atomic samples correlated by QND measurements of the atomic state population by means of 
optical phase shifts. 

We generalized the interferometric set-up introduced in Ref.~\cite{noi3}
to measure the field phase shift, by adding a photo-detector surrounding the two atomic clouds
to allow detection of the scattered photons. In this way we were able to study
the effects of spontaneous scattering, neglected in the ideal treatment \cite{noi3}.
We have introduced a formalism of conditional quantum dynamics in terms of reset operators that, 
acting on the initial atomic wave functions, yield the
new state vector modified according to the result of the photo-detection of the spontaneously scattered photons.
We have then determined the associated effective Hamiltonian that provides the time evolution of 
the system in the instances that the photons reach the detector measuring the phase shifts.
We have finally implemented the formalism in numerical simulations, 
by means of which the evolution of the entanglement of the two atomic samples has been monitored. 
We have compared the results of the simulations in the presence of spontaneous 
scattering with the ideal evolution that neglects this effect. 
Under the assumptions and approximations adopted, we have shown that the 
effects of the spontaneous scattering are rather small, and thus that the
QND interferometric scheme should be robust in realistic situations.

The model dynamics introduced in the present work 
should be useful also for further studies on the dynamics of entanglement 
of atomic samples. In particular, it would be very interesting to combine the entangling 
protocol presented here with some feedback scheme \cite{thomsen02,BerrySanders2}, in order 
to drive the system with very high probability toward the maximally entangled state. 
Moreover, the conditional dynamics could be used to investigate the entanglement properties 
of the state space in which the atomic system state vector is projected by the measurement, 
with the scope of exploiting these properties to realize more efficient entangling protocols.

\end{document}